\documentclass[a4paper,fleqn]{cas-sc}

\usepackage[numbers]{natbib}
\usepackage{wrapfig}
\usepackage{gensymb}
\def\tsc#1{\csdef{#1}{\textsc{\lowercase{#1}}\xspace}}
\tsc{WGM}
\tsc{QE}
\tsc{EP}
\tsc{PMS}
\tsc{BEC}
\tsc{DE}

\begin{document}
\let\WriteBookmarks\relax
\def\floatpagepagefraction{1}
\def\textpagefraction{.001}
\shorttitle{Leveraging social media news}
\shortauthors{P. Singh et~al.}

\title [mode = title]{Steady-State Model of VSC based FACTS Devices using Flexible Holomorphic Embedding: (SSSC and IPFC)}                      



\author[1]{Pradeep Singh}[type=editor,
                        auid=000,bioid=1,
                        prefix=,
                        role=,
                        orcid=]
\ead{psn121988@gmail.com}


\address[]{Department of Electrical Engineering, Indian Institute of Technology, Delhi,
30332 INDIA}


\author[2]{Nilanjan Senroy}[%
   role=,
   suffix=,
   ]
\ead{nsenroy@ee.iitd.ac.in}







\begin{abstract}
For proper planning, operation, control and protection of the power system, the development of suitable steady-state mathematical model of FACTS devices is a key issue. The Fast and Flexible Holomorphic Embedding (FFHE) method converges faster and provides the flexibility to use any state as an initial guess. But to investigate the effect and ability of FACTS devices using FFHE technique, it is necessary to develop an embedded system for these devices. Therefore, this paper presents FFHE-based embedded system for VSC-based FACTS controllers, such as, SSSC and IPFC. The embedded system is also proposed for their controlling modes. 

The introduced embedded system is flexible which allows to take any state as an initial guess instead of fixed state, which leads towards the reduced runtime and decrease the required number of terms, as compared to standard HELM. To demonstrate the effectiveness and practicability, the proposed FFHE-based models of FACTS devices have been tested for several cases. Further, the developed recursive formulas for power balance equations, devices' physical constraints and their controlling modes are thoroughly investigated and examined. From several tests, it is found that the proposed FFHE-based FACTS models requires less execution time and reduce the error at higher rate.
\end{abstract}




\maketitle

\section{Introduction}

Load-flow analysis is one of the fundamental approach of numerical analysis in electric power systems. It provides information regarding the steady-state magnitude and phase angle of voltage at each bus  which can be used in power system planning, operation, protection and control \cite{stott1974review}. The outcomes of the load-flow analysis serve as the starting point in various studies such as dynamic simulation, stability assessment, state-estimation, transfer capability calculation, control and protection schemes of the power system. The conventional load-flow methods are based on numerical iterative techniques such as Gauss-Seidal (GS), Newton-Raphson (NR), Fast Decoupled (FD), and their variants. However, these iterative techniques are primarily constrained by requirement of a good initial guess and they offer slow convergence or even divergence in some cases. It is obvious that the aforesaid problems are further aggravated by the integration of power electronics based control devices and renewable energy \cite{zhang2003advanced, zhang2003modelling, wei2004common}. These traditional iterative load-flow methods can diverge mainly in two sense: (i) solution exists but numerical load-flow method diverges due to bad initial guess, (ii) solution doesn't exists but load-flow method converges to a spurious solution \textit{i.e.} unable to detect the non-existence. To circumvent these problems, a Holomorphic Embedding Load-Flow Method (HELM) has been developed in \cite{trias2012holomorphic}. 

The considerable work on modeling of slack bus, load bus, and generator bus along-with reactive power limit constraints of generator bus have been found in \cite{trias2012holomorphic, rao2016holomorphic, rao2017estimating, wallace2016alternative, subramanian2013pv, trias2016holomorphic, liu2017real}; and other significant advancements in the area of HELM can be found in \cite{yao2018voltage, rao2019three, yao2019efficient, wu2019holomorphic, asl2019holomorphic}. The holomorphic embedded equations have been obtained by separating the shunt and series part of the admittance matrix; and initially the system is assumed to be without load and generation except the slack bus. The performance of this method has also been investigated for calculating the available voltage stability margin and it is found that HELM performs at par with the continuation power-flow method \cite{liu2017online, liu2017real, singh2020extended, wang2017multi}. The key benefits of HELM are deterministic initial guess and guarantee of convergence if the solution exists, but, at the cost of more execution time than NR method \cite{sauter2017comparison, freitas2019restarted, li2018numerical}.  HELM is 10 to 20 times slower than NR method \cite{trias2016holomorphic}. Therefore, to address this problem, a Fast, Flexible Holomorphic Embedded (FFHE) method has been proposed in \cite{chiang2017novel}. This developed embedded system is termed as flexible because this method allows to set any state as an initial guess. In most cases, FFHE is fast because it takes less number of terms to converge if initial guesses are promising.

The inherent benefits of HELM encouraged the development of holomorphic embedded models of various Flexible Alternating Current Transmission System (FACTS) controllers to improve the overall performance of the power system by their optimal integration \cite{wei2004common, jiang2008novel, singh2017amalgam}. Thyristor-based FACTS controllers are represented as a variable admittance or impedance in \cite{basiri2017holomorphic}. However, this approach does not accurately exhibit the characteristics of second-generation (\textit{i.e.} Voltage Source Converter (VSC)-based) FACTS controllers. A more accurate model of VSC-based FACTS controllers involves representing it as a variable voltage or current source \cite{jiang2008novel, singh2017amalgam}. The VSC-based FACTS controllers are preferred because they inject less harmonics, provide independent control, fast dynamic response and higher flexibility as compared to Thyristor-based FACTS controller.

In the sphere of HELM, no research work has been found on modeling of Voltage Source Converter (VSC)-based FACTS devices using Holomorphic Embedding (HE) except modeling of Static Synchronous Compensator (STATCOM) \cite{singh2019statcom}. Therefore, the main aim of this work is to develop the FFHE based models of VSC-based FACTS controllers. This paper proposes the FFHE based models of two VSC-based series FACTS devices, namely Static Synchronous Series Compensator (SSSC) and Interline Power Flow Controller (IPFC).


The remaining part of this paper is organised as follows: Section \ref{PM} deals with the methodology used to develop the FFHE based models of SSSC and IPFC along-with their controlling modes. In Section \ref{results}, the key findings and numerical results of the introduced embedded system have been discussed. Section \ref{conclusions}, presents the conclusion of this study.

\section{Proposed Holomorphic Embedding Formulations} \label{PM}
In this section, a new embedded system for VSC-based FACTS controllers such as SSSC and IPFC are developed and the numerical values of all unknown variables along-with their operation bounds are investigated. To develop the steady-state FFHE based HE models of the FACTS controllers, it is assumed that the stated devices and system are three-phase balanced and the harmonics generated by them are negligible. Firstly, the holomorphic embedded system and finally recursive relationships for the stated devices have been developed. 

\subsection{SSSC Modeling} \label{SSSC_modeling}
\begin{figure}
	\centering
		\includegraphics[scale=1]{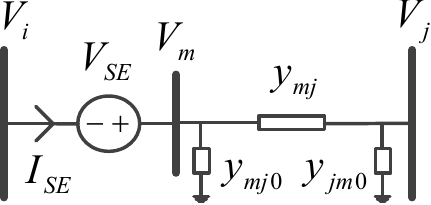}
	\caption{Equivalent circuit of a SSSC.}
	\label{SSSC_Equivalent}
\end{figure}

A SSSC is a series connected FACTS device, and it can generate or inject a series voltage into the transmission line, which can be regulated to change the reactance of the transmission line. So, the active and reactive power-flows through the line or the magnitude of bus voltage, where the SSSC is connected can be controlled. The SSSC consists of a coupling transformer, a AC/DC voltage source converter and a capacitor. A schematic equivalent circuit diagram of a SSSC is shown in Figure \ref{SSSC_Equivalent}, where it is assumed that the transmission line is series connected via the extra bus $m$. In other words, a series FACTS device is modelled by assuming a separate branch in the network and this branch is not considered in the admittance matrix.

In Figure \ref{SSSC_Equivalent}, $y_{mj}=y_{ij}+y_{SE}$ is the admittance between bus $m$ and bus $j$; $y_{SE}$ is the admittance of the series coupling transformer, $V_{SE}$ is the injected series voltage phasor, and $I_{SE}$ is the current flowing through SSSC.
From Figure \ref{SSSC_Equivalent}, the power balance equation (PBE) for buses $i$ and $m$ can be expressed as follows:
\begin{eqnarray}
V^{*}_{i}\sum_{k=1}^{N} Y_{ik}V_{k} + V^{*}_{i}I_{SE}= S^{*}_{i}
\label{SSSC_Node_S}\\
V^{*}_{m}\sum_{k=1}^{N} Y_{mk}V_{k} - V^{*}_{m}I_{SE}= S^{*}_{m}
\label{SSSC_Node_R}
\end{eqnarray}
where, $'*'$ represents conjugation, $V_{i}$ and $Y_{ik}$ are the voltage phasor of bus $i$ and $(i,k)^{th}$ entry of the bus admittance matrix. $S_{i}=(P_{Gi}+jQ_{Gi})-(P_{Li}+jQ_{Li})$ is the injected complex power at bus $i$. $P_{Gi}$ and $Q_{Gi}$ are the active and reactive power generation, where as $P_{Li}$ and $Q_{Li}$ are active and reactive power load at bus $i$ respectively. The symbols $N$, $N_{PV}$, and $N_{PQ}$ denotes the set of total, generator and load buses respectively.

Now, to establish the recurrence relationship for (\ref{SSSC_Node_S}) and (\ref{SSSC_Node_R}), these equations must be embedded with a complex parameter $\alpha$ in such a way that the resultant equations should satisfy the requirement of embedding \textit{i.e.} at $\alpha =1$, the holomorphic embedding formulation must be equal to the original equation and at $\alpha =0$, the solution of the system is already known or can be obtained trivially. In  (\ref{SSSC_Node_S}) and (\ref{SSSC_Node_R}), the current through SSSC is treated as a free variable function of $\alpha$ and the following new embedded system is introduced for (\ref{SSSC_Node_S}) and (\ref{SSSC_Node_R}):
\begin{equation}
V^{*}_{i}(\alpha^{*})\sum_{k=1}^{N} Y_{ik}V_{k}(\alpha) + V^{*}_{i}(\alpha^{*})I_{SE}(\alpha)= C^{*}_{i}\sum_{k=1}^{N} Y_{ik}C_{k}+C^{*}_{i}D_{SE} + \alpha \Bigg [S^{*}_{i} - C^{*}_{i}\sum_{k=1}^{N} Y_{ik}C_{k}-C^{*}_{i}D_{SE} \Bigg ]
\label{SSSC_Node_S_HE}
\end{equation}
\begin{equation}
V^{*}_{m}(\alpha^{*})\sum_{k=1}^{N} Y_{mk}V_{k}(\alpha) - V^{*}_{m}(\alpha^{*})I_{SE}(\alpha)= C^{*}_{m}\sum_{k=1}^{N} Y_{mk}C_{k}
-C^{*}_{m}D_{SE} + \alpha \Bigg [S^{*}_{m} - C^{*}_{m}\sum_{k=1}^{N} Y_{mk}C_{k}+C^{*}_{m}D_{SE} \Bigg ]
\label{SSSC_Node_R_HE}
\end{equation}

Note that the equation (\ref{SSSC_Node_S_HE}) and (\ref{SSSC_Node_R_HE}) satisfy the requirement of embedding at reference state $\alpha_{0}=0$ and the target state $\alpha_{1}=1$. The constant $C_{k}\in \mathbb{C}\setminus \{0\}$ and $D_{SE}\in \mathbb{C}\setminus \{0\}$ are adjustable and can be of any pre-specified values. The constant $C_{k}$ and $D_{SE}$ are used to represent the initial value of voltages and SSSC's current respectively. At $\alpha = 0$, the solution of (\ref{SSSC_Node_S_HE}) and (\ref{SSSC_Node_R_HE}) gives $V_{k}[0] = C_{k}$ and $I_{SE}[0]=D_{SE}$.

The integration of an SSSC into transmission line introduces a complex variable $I_{SE}=I_{SEre}+jI_{SEim}$ into power balance equation, which adds two unknown variables (\textit{i.e. $I_{SEre}$ and $I_{SEim}$}). Therefore, two supplementary equations are required to find out the unique solution of unknown variables. These supplementary equations are related to the SSSC real power injection constraint and reference quantity to be controlled by the SSSC. In steady-state, the ideal SSSC can only inject the reactive power but can't inject real power into AC system unless there is a connected energy source. Mathematically, from Figure \ref{SSSC_Equivalent} it can be expressed as:
\begin{equation}
\Re \big [ (V_{m}-V_{i})I^{*}_{SE} \big ] = 0
\label{SSSC_Active_Cons}
\end{equation} 
$\Re(\bullet)$ and $\Im(\bullet)$ are the real and imaginary operators. Equation (\ref{SSSC_Active_Cons}) can be embedded as follows:
\begin{equation}
\Re \bigg [ \big \{V_{m}(\alpha)-V_{i}(\alpha)\big \}I^{*}_{SE}(\alpha^{*}) \bigg ] = \Re \big [ \{C_{m}-C_{i}\}D^{*}_{SE} \big ]
 - \alpha \Re \big [ \{C_{m}-C_{i}\}D^{*}_{SE} \big ]
\label{SSSC_Active_Cons_HE}
\end{equation}
It has been recognised that the SSSC has the ability to control various parameters but only one variable can be controlled independently as the degree of freedom of SSSC is one \cite{zhang2003advanced}. The proposed model of SSSC has been tested with the following control modes.
\subsubsection{Case 1: Active power-flow control}
In this case, the injected series voltage $V_{SE}$ is regulated in such a manner that the specified active power-flow through transmission line can be achieved. Mathematically, it can be expressed as follows:
\begin{equation}
\Re \big [ V_{i}I^{*}_{SE} \big ] = P^{SP}_{im}
\label{Mode1}
\end{equation}
where, $P^{SP}_{im}$ is the specified real power-flow between bus $i$ and $m$. The proposed embedding for (\ref{Mode1}) is as follows:
\begin{equation}
\Re \big [ V_{i}(\alpha) I^{*}_{SE}(\alpha^{*}) \big ] = \Re (C_{i}D^{*}_{SE})+ \alpha \big [ P^{SP}_{im} - \Re (C_{i}D^{*}_{SE} ) \big ]
\label{Mode1_HE}
\end{equation}

\subsubsection{Case 2: Reactive power-flow control}
In this case, the SSSC is used to control transmission line reactive power-flow to a specified reactive power reference by adjusting the SSSC injected voltage phasor. Mathematically, it can be written as follows:
\begin{equation}
\Im \big [ V_{i}I^{*}_{SE} \big ] = Q^{SP}_{im}
\label{Mode2}
\end{equation}
where, $Q^{SP}_{im}$ is the specified reactive power-flow between bus $i$ and $m$. The proposed embedding for (\ref{Mode2}) is as follows:
\begin{equation}
\Im \big [V_{i}(\alpha) I^{*}_{SE}(\alpha^{*})\big ] = \Im (C_{i}D^{*}_{SE})+ \alpha \big [ Q^{SP}_{im} - \Im (C_{i}D^{*}_{SE} ) \big ]
\label{Mode2_HE}
\end{equation}

\subsubsection{Case 3:  Reactive power injection control}
In this case, reactive power injected by the SSSC is regulated to the target value of reactive power injection by adjusting the SSSC injected voltage phasor and it can be expressed as follows:
\begin{equation}
\Im \big [ (V_{m}-V_{i})I^{*}_{SE} \big ] = Q^{SP}_{SE}
\label{Mode3}
\end{equation}
where, $Q^{SP}_{SE}$ is the specified reactive power injected by the SSSC. The proposed embedding for (\ref{Mode3}) is as follows:
\begin{equation}
\Im \bigg [ \big \{V_{m}(\alpha)-V_{i}(\alpha) \big \}I^{*}_{SE}(\alpha^{*}) \bigg ] = \Im \big [ (C_{m}-C_{i})D^{*}_{SE} \big ] + \alpha \bigg [ Q^{SP}_{im} - \Im \big \{ (C_{m}-C_{i})D^{*}_{SE} \big \} \bigg ]
\label{Mode3_HE}
\end{equation}

\subsubsection{Case 4: Bus voltage magnitude control}
The local or remote bus voltage magnitude is controlled to a specified reference value by regulating the SSSC output voltage phasor. Mathematically, the bus voltage magnitude constraint for bus $i$ can be expressed as follows:
\begin{equation}
|V_{i}| = V^{SP}_{i}
\label{Mode4}
\end{equation}
where, $V^{SP}_{i}$ is the specified value of $i^{th}$ bus voltage magnitude. The mathematical expression of this control mode is identical to PV bus voltage magnitude constraint expression, therefore, it is embedded in the similar fashion as in \cite{chiang2017novel} and it can be expressed as follows:
\begin{equation}
V_{i}(\alpha)V^{*}_{i}(\alpha^{*}) = C_{i}C^{*}_{i} + \alpha \big [ (V^{SP}_{i})^2 - C_{i}C^{*}_{i} \big ]
\label{Mode4_HE}
\end{equation}  

\subsubsection{Case 5: Control of SSSC's injected voltage magnitude}
SSSC can be used to regulate the magnitude of $V_{SE}$ to a specified reference and this control constraint can be expressed as follows: 
\begin{equation}
|V_{SE}| = V^{SP}_{SE}
\end{equation}
where, $V^{SP}_{SE}$ is the specified value of injected voltage. The SSSC's injected voltage can be expressed in terms of reactive power injected by the source as follows:
\begin{equation}
\Im \bigg [ \frac{(V_{m}-V_{i})|I_{SE}|}{I_{SE}} \bigg ] = V^{SP}_{SE}
\label{Mode5}
\end{equation}
The proposed embedding for (\ref{Mode5}) can be expressed as follows:
\begin{equation}
\Im \big [(V_{m}(\alpha)-V_{i}(\alpha))|I_{SE}(\alpha)|F_{SE}(\alpha) \big ] = \Im \bigg \{\frac{(C_{m}-C_{i})|D_{SE}|}{D_{SE}} \bigg \} + \alpha \bigg [V^{SP}_{SE}-\Im \bigg \{\frac{(C_{m}-C_{i})|D_{SE}|}{D_{SE}} \bigg \} \bigg ] 
\label{Mode5_HE}
\end{equation}
where, $F_{SE}(\alpha)$ is the reciprocal of $I_{SE}(\alpha)$. 

\subsubsection{Case 6: Control of equivalent series reactance} 
A SSSC can also be viewed as a equivalent imaginary impedance. In this case, SSSC's injected voltage phasor is adjusted to control the equivalent imaginary impedance to a specified impedance reference value. Mathematically, such a constraint can be written as follows:   
\begin{equation}
\Im \Bigg [ \frac{V_{m}-V_{i}}{I_{SE}} \Bigg ]= X^{SP}_{eq(SE)}
\label{Mode6}
\end{equation}
where, $X^{SP}_{eq(SE)}$ is the specified equivalent imaginary reactance. The proposed embedding for (\ref{Mode6}) is as follows:
\begin{equation}
\Im \Bigg [ \frac{V_{m}(\alpha)-V_{i}(\alpha)}{I_{SE}(\alpha)} \Bigg ] =  \alpha \Bigg [ X^{SP}_{eq(SE)} - \Im \Bigg ( \frac{C_{m}-C_{i}}{D_{SE}} \Bigg ) \Bigg ]
+ \Im \Bigg ( \frac{C_{m}-C_{i}}{D_{SE}} \Bigg )
\label{Mode6_HE}
\end{equation}

The system of equations (\ref{SSSC_Node_S_HE}), (\ref{SSSC_Node_R_HE}), (\ref{SSSC_Active_Cons_HE}) and any one equation from the set of (\ref{Mode1_HE}), (\ref{Mode2_HE}), (\ref{Mode3_HE}), (\ref{Mode4_HE}), (\ref{Mode5_HE}), and (\ref{Mode6_HE}) can describe the behaviour of SSSC. 

Generally, a system of linear equations is derived to evaluate the unknown variables (\textit{i.e.} $V_{kre}$, $V_{kim}$, $I_{SEre}$ and $I_{SEim}$) and the coefficient matrix which has the entries of products and sums. The formulation of the coefficient is more complex and less efficient as compared to the coefficient matrix formulation in basic HELM because the later one doesn't involve the products of coefficient of voltages and currents. This shortcoming is outweighed by the reduced number of steps required to evaluate unknown variables. In other words, it is faster than the basic HELM. 
\begin{figure*}[b]
\hrulefill
\begin{equation}
\resizebox{0.95\hsize}{!}{%
\rotatebox{0}{$ [A^{SC}] =\begin{bmatrix} \begin{array}{cccccccccccc}   
1  	 &   0 	  & 0 &  0 & 0 & 0 & 0 & 0 & 0  & 0 & 0 & 0  \\
0  	 &   1    & 0 &  0 & 0 & 0 & 0 & 0 & 0  & 0 & 0 & 0  \\

\mu_{\mathcal{GF}} & \xi_{\mathcal{GF}} & \mu_{\mathcal{GG}} & \xi_{\mathcal{GG}} &  \mu_{\mathcal{GL}} & \xi_{\mathcal{GL}} &   \mu_{\mathcal{G}i} & \xi_{\mathcal{G}i} & \mu_{\mathcal{G}m} & \xi_{\mathcal{G}m}  & 0  & 0\\

 0 & 0 & C_{\mathcal{G}re} &  C_{\mathcal{G}im} & 0 & 0 &0 & 0 & 0  & 0  & 0 & 0	  \\

\mu_{\mathcal{LF}} & \xi_{\mathcal{LF}} & \mu_{\mathcal{LG}} & \xi_{\mathcal{LG}} &  \mu_{\mathcal{LL}} & \xi_{\mathcal{LL}} &  \mu_{\mathcal{L}i}& \xi_{\mathcal{L}i} & \mu_{\mathcal{L}m} & \xi_{\mathcal{L}m} & 0  & 0\\

 - \xi_{\mathcal{LF}} & \mu_{\mathcal{LF}} & -\xi_{\mathcal{LG}} & \mu_{\mathcal{LG}} &  \mu^{\bigstar}_{\mathcal{LL}} & \xi^{\bigstar}_{\mathcal{LL}} & - \xi_{\mathcal{L}i} & \mu_{\mathcal{L}i} & - \xi_{\mathcal{L}m} & \mu_{\mathcal{L}m} & 0 & 0\\

\mu_{i\mathcal{F}} & \xi_{i\mathcal{F}} & \mu_{i\mathcal{G}} & \xi_{i\mathcal{G}} &  \mu_{i\mathcal{L}}  & \xi_{i\mathcal{L}} &  \mu_{ii} + D_{SEre} & \xi_{ii}+ D_{SEim}  & \mu_{im} & \xi_{im} & C_{ire}  & C_{iim}\\

- \xi_{i\mathcal{F}} & \mu_{i\mathcal{F}} & - \xi_{i\mathcal{G}} & \mu_{i\mathcal{G}} & - \xi_{i\mathcal{L}} & \mu_{i\mathcal{L}} &  \mu^{\bigstar}_{ii} + D_{SEim} & \xi^{\bigstar}_{ii}- D_{SEre}  & -\xi_{im} & \mu_{im} & -C_{iim}  & C_{ire}\\

\mu_{m\mathcal{F}} & \xi_{m\mathcal{F}} & \mu_{m\mathcal{G}} & \xi_{m\mathcal{G}} &  \mu_{m\mathcal{L}}  & \xi_{m\mathcal{L}} &  \mu_{mi} &  \xi_{mi} & \mu_{mm} - D_{SEre} & \xi_{mm} - D_{SEim} & - C_{mre}  & - C_{mim}\\

- \xi_{m\mathcal{F}} & \mu_{m\mathcal{F}} & - \xi_{m\mathcal{G}} & \mu_{m\mathcal{G}} & - \xi_{m\mathcal{L}} & \mu_{m\mathcal{L}} &  -\xi_{mi} &  \mu_{mi} & \mu^{\bigstar}_{mm} - D_{SEim} & \xi^\bigstar_{mm} + D_{SEim} &  C_{mim}  & - C_{mre}\\

0 & 0 & 0 & 0 & 0 & 0 & - D_{SEre} & - D_{SEim} & D_{SEre} & D_{SEim} & C_{mre} - C_{ire} & C_{mim} - C_{mim}\\

\multicolumn{12}{c}{ \cdots Entries~related~to~selected~control~mode~from~(\ref{SSSC_Modes}) \cdots} \\
\end{array} 
\end{bmatrix}$}}
\label{SSSC_Coeff_Matrix}
\end{equation}
\begin{equation}
\resizebox{0.94\hsize}{!}{%
\rotatebox{0}{$\begin{bmatrix} \begin{array}{cccccccccccc}   

 0  & 0 	 &   0 & 0 &  0 & 0 & D_{SEre}  & D_{SEim} & 0 &  0 & C_{ire} & C_{iim} \\
 0  & 0      &   0 & 0 &  0 & 0 & -D_{SEim}  & D_{SEre} & 0 &  0 & C_{iim}& -C_{ire} \\
 0  & 0      &   0 & 0 &  0 & 0 & D_{SEim}  & -D_{SEre} & -D_{SEim} &  D_{SEre} & C_{mim} -C_{iim} & C_{mre} -C_{ire} \\
 0  & 0      &   0 & 0 &  0 & 0 & C_{ire}  & C_{iim} & 0 &  0 & 0 & 0 \\
 0  & 0      &   0 & 0 &  0 & 0 & -\Im \Big(\frac{|D_{SE}|}{D_{SE}}\Big)  & -\Re \Big(\frac{|D_{SE}|}{D_{SE}}\Big) & \Im \Big(\frac{|D_{SE}|}{D_{SE}}\Big) &  \Re \Big(\frac{|D_{SE}|}{D_{SE}}\Big) &  \Im \Big (\frac{2D_{SEre}(C_{m}-C_{i})}{2D_{SE}|D_{SE}|}\Big ) - \Im \Big (\frac{(C_{m}-C_{i})|D_{SE}|}{D^{2}_{SE}}\Big )  &  \Im \Big (\frac{2D_{SEim}(C_{m}-C_{i})}{2D_{SE}|D_{SE}|}\Big ) - \Re \Big (\frac{(C_{m}-C_{i})|D_{SE}|}{D^{2}_{SE}}\Big )  \\
 0  & 0      &   0 & 0 &  0 & 0 & \frac{D_{SEim}}{|D|^{2}}  & \frac{-D_{SEre}}{|D|^{2}} & \frac{-D_{SEim}}{|D|^{2}} &  \frac{D_{SEre}}{|D|^{2}} & \Im\Big(\frac{C_{i}-C_{m}}{D^{2}_{SE}}\Big ) & \Re\Big(\frac{C_{i}-C_{m}}{D^{2}_{SE}}\Big ) \\
\end{array} 
\end{bmatrix} \begin{bmatrix}
       X^{SC}
        \end{bmatrix} =~\begin{bmatrix} \begin{array}{c}
        		   \Gamma^{SC}_{M1}  \\
        		   \Gamma^{SC}_{M2}  \\
        		   \Gamma^{SC}_{M3}  \\
                   \Gamma^{SC}_{M4}  \\
                   \Gamma^{SC}_{M5}  \\
 				\end{array} 
                \end{bmatrix}$}}
\label{SSSC_Modes} 
\end{equation}
\begin{equation}
\resizebox{0.9\hsize}{!}{%
\rotatebox{0}{$[X^{SC}] = \begin{bmatrix} \begin{array}{cccccccccccc}
V_{\mathcal{F}re}[n] & V_{\mathcal{F}im}[n] & V_{\mathcal{G}re}[n] & V_{\mathcal{G}im}[n] & V_{\mathcal{L}re}[n] & V_{\mathcal{L}im}[n] & V_{ire}[n] & V_{iim}[n] & V_{mre}[n] & V_{mim}[n] & I_{SEre}[n] & I_{SEim}[n]
\end{array}\end{bmatrix}^{'} $}}
\label{SSSC_Vector} 
\end{equation}
\begin{equation}
\resizebox{0.9\hsize}{!}{%
\rotatebox{0}{$[B^{SC}] = \begin{bmatrix} \begin{array}{cccccccccccc}
\Re[\Gamma_{\mathcal{F}}] & \Im[\Gamma_{\mathcal{F}}] & \Gamma_{\mathcal{G}} & \Gamma_{\mathcal{GV}}  & \Re[\Gamma_{\mathcal{L}}] & \Im[\Gamma_{\mathcal{L}}] & \Re[\Gamma^{SC}_{i}] & \Im[\Gamma^{SC}_{i}] & \Re[\Gamma^{SC}_{m}] & \Im[\Gamma^{SC}_{m}] & \Gamma^{SC}_{PBE}         & \Gamma^{SC}_{Mi}     	  
\end{array}\end{bmatrix}^{'} $}} 
\label{SSSC_RHS}
\end{equation}
\end{figure*}

To obtain the system of linear equations, all known and unknown variables are moved to RHS and LHS respectively. For this purpose, variables are separated into real and imaginary parts as follows: $Y_{ik}=G_{ik}+jB_{ik}$, $V_{k}=V_{kre}+jV_{kim}$ and $I_{SE}=I_{SEre}+jI_{SEim}$. The system of linear equations can be viewed as $[A^{SC}]_{(\Upsilon\times\Upsilon)}[X^{SC}]_{(\Upsilon\times 1)}=[B^{SC}]_{(\Upsilon\times 1)}$, where $\Upsilon=2(N+n_{SC})$ and $n_{SC}$ is the number of SSSCs connected into the system. The coefficient matrix $[A^{SC}]$, unknown vector $[X^{SC}]$ and known vector $[B^{SC}]$ when SSSC is incorporated in the system are given by (\ref{SSSC_Coeff_Matrix}), (\ref{SSSC_Vector}) and (\ref{SSSC_RHS}) respectively (shown at the bottom of this page). Here, subscripts $\mathcal{F}$, $\mathcal{G}\in N_{PV}$, $\mathcal{L}\in N_{PQ}$, $i$, and $m$ denote the slack bus, PV bus, load bus, SSSC sending end and receiving end connected bus respectively. Equation (\ref{SSSC_Modes}) represents the different control modes of SSSC in linear matrix form. Note that for calculating the $n^{th}$ coefficients of the unknown vector, the maximum required degree of known vector is $n-1$, but for representation purpose $n-1$ index are omitted from known vector throughout this paper. The holomorphic embedded system for slack bus, generator buses and load buses are adopted from \cite{chiang2017novel}. At $\alpha=1$, the numerical values of unknown variables are evaluated using determinant method as discussed in \cite{chiang2017novel}. The entries of coefficient matrix and known vector are given in Appendix A.

\subsection{IPFC Modeling}
An IPFC may be constructed by combining two or more series connected converters, which extends the philosophy of power-flow control beyond what is achievable with one converter series-connected FACTS device \textit{i.e.} SSSC. IPFC can enhance power transfer capability of transmission line and maximize the use of existing transmission infrastructure. 

\begin{figure}[!h]
	\centering
		\includegraphics[scale=1.3]{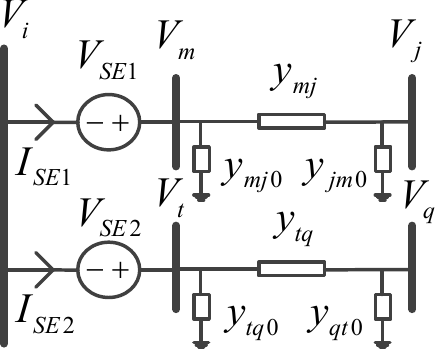}
	\caption{Equivalent circuit of a IPFC.}
	\label{IPFC_Equivalent}
\end{figure}
The simplest IPFC can be configured with two series connected converters with two transmission lines via series coupling transformers. Figure \ref{IPFC_Equivalent} shows the equivalent circuit of IPFC with two controllable voltage source. Note that the magnitude of series injected voltages are independent of the magnitude of the bus voltages because it is injected by voltage source converters. So, the controllable parameter of each IPFC branch is series injected voltage phasor. Therefore, other parameters (\textit{e.g.} line active power-flow, line reactive power-flow \textit{etc.}) of the system can be controlled by regulating the series injected voltage phasor. The injected reactive power into system is provided by the VSC, but injection of real power into the system has to be provided by other voltage source converters via a common DC link. Therefore, an IPFC with $n_{IP}$ branches can independently control $2n_{IP}-1$ parameters as one parameters is responsible for real power balance of the device. Therefore, this configuration can simultaneously control three quantities of power system independently. The detailed descriptions of the IPFC and its features may be found in \cite{zhang2006novel, vinkovic2009current}. 

From Figure \ref{IPFC_Equivalent}, the PBE for buses $i$, $m$ and $t$ can be expressed as follows:
\begin{equation}
V^{*}_{i}\sum_{k=1}^{N} Y_{ik}V_{k} + V^{*}_{i}I_{SE1} + V^{*}_{i}I_{SE2} = S^{*}_{i}
\label{IPFC_Node_S} 
\end{equation}
\begin{equation}
V^{*}_{m}\sum_{k=1}^{N} Y_{mk}V_{k} - V^{*}_{m}I_{SE1}= S^{*}_{m}
\label{IPFC_Node_R1} 
\end{equation}
\begin{equation}
V^{*}_{t}\sum_{k=1}^{N} Y_{tk}V_{k} - V^{*}_{t}I_{SE2}= S^{*}_{t}
\label{IPFC_Node_R2}
\end{equation}
The active power can be exchanged amongst the converters via a common DC link, but the sum of active power exchanged must be equal to zero. Mathematically, it can be expressed as follows:
\begin{equation}
\Re \big [ (V_{m}-V_{i})I^{*}_{SE1} \big ] + \Re \big [ (V_{t}-V_{i})I^{*}_{SE2} \big ]= 0
\label{IPFC_Active_Cons}
\end{equation} 
The following new embedded system is introduced for (\ref{IPFC_Node_S})-(\ref{IPFC_Active_Cons}):

\begin{multline}
V^{*}_{i}(\alpha^{*})\sum_{k=1}^{N} Y_{ik}V_{k}(\alpha) + V^{*}_{i}(\alpha^{*}) \big \{ I_{SE1}(\alpha)  + I_{SE2}(\alpha)\big \}= C^{*}_{i}\sum_{k=1}^{N} Y_{ik}C_{k}
+ C^{*}_{i} \big (D_{SE1} + D_{SE2}\big ) + \alpha \Bigg [S^{*}_{i} - C^{*}_{i}\sum_{k=1}^{N} Y_{ik}C_{k}
\\-C^{*}_{i}\big (D_{SE1} + D_{SE2}\big )\Bigg ]
\label{IPFC_Node_S_HE}
\end{multline}
\begin{equation}
V^{*}_{m}(\alpha^{*})\sum_{k=1}^{N} Y_{mk}V_{k}(\alpha) - V^{*}_{m}(\alpha^{*})I_{SE1}(\alpha)= C^{*}_{m}\sum_{k=1}^{N} Y_{mk}C_{k}
-C^{*}_{m}D_{SE1} + \alpha \Bigg [S^{*}_{m} - C^{*}_{m}\sum_{k=1}^{N} Y_{mk}C_{k}+C^{*}_{m}D_{SE1} \Bigg ]
\label{IPFC_Node_R1_HE}
\end{equation}
\begin{equation}
V^{*}_{t}(\alpha^{*})\sum_{k=1}^{N} Y_{tk}V_{k}(\alpha) - V^{*}_{t}(\alpha^{*})I_{SE2}(\alpha)= C^{*}_{t}\sum_{k=1}^{N} Y_{tk}C_{k}
\\-C^{*}_{t}D_{SE2} + \alpha \Bigg [S^{*}_{t} - C^{*}_{t}\sum_{k=1}^{N} Y_{tk}C_{k}+C^{*}_{t}D_{SE2} \Bigg ]
\label{IPFC_Node_R2_HE}
\end{equation}
\begin{multline}
\Re \bigg [ \big \{V_{m}(\alpha)-V_{i}(\alpha)\big \}I^{*}_{SE1}(\alpha^{*}) + \big \{V_{t}(\alpha)-V_{i}(\alpha)\big \}I^{*}_{SE2}(\alpha^{*})\bigg ] 
= \Re \big [ (C_{m}-C_{i})D^{*}_{SE1} + (C_{t}-C_{i})D^{*}_{SE2}\big ]
 \\- \alpha \Re \big [ (C_{m}-C_{i})D^{*}_{SE1} + (C_{t}-C_{i})D^{*}_{SE2}\big ]
\label{IPFC_Active_Cons_HE}
\end{multline}
\begin{figure*}[!b]
\hrulefill
\begin{equation}
\resizebox{1\hsize}{!}{%
\rotatebox{0}{$ [A^{IP}] = \begin{bmatrix} \begin{array}{cccccccccccccccc}   
1  	 &   0 	  & 0 &  0 & 0 & 0 & 0 & 0 & 0  & 0 & 0 & 0  & 0 & 0 & 0 & 0\\
0  	 &   1    & 0 &  0 & 0 & 0 & 0 & 0 & 0  & 0 & 0 & 0  & 0 & 0 & 0 & 0\\

\mu_{\mathcal{GF}} & \xi_{\mathcal{GF}} & \mu_{\mathcal{GG}} & \xi_{\mathcal{GG}} &  \mu_{\mathcal{GL}} & \xi_{\mathcal{GL}} &   \mu_{\mathcal{G}i} & \xi_{\mathcal{G}i} & \mu_{\mathcal{G}m} & \xi_{\mathcal{G}t}  & \mu_{\mathcal{G}t} & \xi_{\mathcal{G}t} & 0  & 0 & 0 & 0\\

 0 & 0 & C_{\mathcal{G}re} &  C_{\mathcal{G}im} & 0 & 0 &0 & 0 & 0  & 0  & 0 & 0 & 0 & 0 & 0 & 0	  \\

\mu_{\mathcal{LF}} & \xi_{\mathcal{LF}} & \mu_{\mathcal{LG}} & \xi_{\mathcal{LG}} &  \mu_{\mathcal{LL}} & \xi_{\mathcal{LL}} &  \mu_{\mathcal{L}i}& \xi_{\mathcal{L}i} & \mu_{\mathcal{L}m} & \xi_{\mathcal{L}m} & \mu_{\mathcal{L}t} & \xi_{\mathcal{L}t} & 0  & 0 & 0  & 0\\

 - \xi_{\mathcal{LF}} & \mu_{\mathcal{LF}} & -\xi_{\mathcal{LG}} & \mu_{\mathcal{LG}} &  \mu^{\bigstar}_{\mathcal{LL}} & \xi^{\bigstar}_{\mathcal{LL}} & - \xi_{\mathcal{L}i} & \mu_{\mathcal{L}i} & - \xi_{\mathcal{L}m} & \mu_{\mathcal{L}m} &          - \xi_{\mathcal{L}t} & \mu_{\mathcal{L}t} & 0 & 0 & 0  & 0\\

\mu_{i\mathcal{F}} & \xi_{i\mathcal{F}} & \mu_{i\mathcal{G}} & \xi_{i\mathcal{G}} &  \mu_{i\mathcal{L}}  & \xi_{i\mathcal{L}} &  \mu_{ii} + D_{SE1re}+ D_{SE2re} & \xi_{ii}+ D_{SE1im}+ D_{SE2im}  & \mu{im} & \xi_{im} & \mu_{it} & \xi_{it} & C_{ire}  & C_{iim}& C_{ire}  & C_{iim}\\

- \xi_{i\mathcal{F}} & \mu_{i\mathcal{F}} & - \xi_{i\mathcal{G}} & \mu_{i\mathcal{G}} & - \xi_{i\mathcal{L}}  & \mu_{i\mathcal{L}} &  \mu^{\bigstar}_{ii} + D_{SE1im}+ D_{SE2im} & \xi^{\bigstar}_{ii}- D_{SE1re}- D_{SE2re}  & -\xi_{im} & \mu_{im}   & -\xi_{it} & \mu_{it} & -C_{iim}  & C_{ire} & -C_{iim}  & C_{ire}\\

\mu_{m\mathcal{F}} & \xi_{m\mathcal{F}} & \mu_{m\mathcal{G}} & \xi_{m\mathcal{G}} &  \mu_{m\mathcal{L}}  & \xi_{m\mathcal{L}} &  \mu_{mi} &  \xi_{mi} & \mu_{mm} - D_{SE1re} & \xi_{mm} - D_{SE1im} & \mu_{mt} & \xi_{mt} & - C_{mre}  & - C_{mim} & 0 & 0\\

- \xi_{m\mathcal{F}} & \mu_{m\mathcal{F}} & - \xi_{m\mathcal{G}} & \mu_{m\mathcal{G}} & - \xi_{m\mathcal{L}}  & \mu_{m\mathcal{L}} &  -\xi_{mi} &  \mu_{mi} & \mu^{\bigstar}_{mm} - D_{SE1im} & \xi^\bigstar_{mm} + D_{SE1im} & -\xi_{mt} & \mu_{mt} &   C_{mim}  & - C_{mre} &  0 & 0\\

\mu_{t\mathcal{F}} & \xi_{t\mathcal{F}} & \mu_{t\mathcal{G}} & \xi_{t\mathcal{G}} &  \mu_{t\mathcal{L}}  & \xi_{t\mathcal{L}} &  \mu_{ti} &  \xi_{ti} & \mu_{tm}  & \xi_{tm}  & \mu_{tt}- D_{SE2re} & \xi_{tt}- D_{SE2im} & 0 & 0 & - C_{tre}  & - C_{tim} \\

- \xi_{t\mathcal{F}} & \mu_{t\mathcal{F}} & - \xi_{t\mathcal{G}} & \mu_{t\mathcal{G}} & - \xi_{t\mathcal{L}} & \mu_{t\mathcal{L}} &  -\xi_{ti} &  \mu_{ti} & -\xi_{tm}   & \mu_{tm}  & \mu^{\bigstar}_{tt}-D_{SE2im} & \xi^{\bigstar}_{tt}+ D_{SE2re} & 0 &   0   &C_{tim} & - C_{mre}  \\

0 & 0 & 0 & 0 & 0 & 0 & - D_{SE1re}- D_{SE2re} & - D_{SE1im}- D_{SE2im} & D_{SE1re} & D_{SE1im} &D_{SE2re} & D_{SE2im}& C_{mre} - C_{ire} & C_{mim} - C_{iim}      & C_{tre} - C_{ire} & C_{tim} - C_{iim}\\

\multicolumn{16}{c}{ \cdots Entries~related~to~selected~control~mode~from~(\ref{SSSC_Modes}) \cdots} \\
\end{array} 
\end{bmatrix} $}}
\label{IPFC_Matrix}
\end{equation}
\begin{equation}
\resizebox{0.9\hsize}{!}{%
\rotatebox{0}{$[X^{IP}] = \begin{bmatrix} \begin{array}{cccccccccccccccc}
V_{\mathcal{F}re}[n]  &
       V_{\mathcal{F}im}[n]  &
       V_{\mathcal{G}re}[n]  &
       V_{\mathcal{G}im}[n]  &
       V_{\mathcal{L}re}[n]  & 
       V_{\mathcal{L}im}[n]  &
       V_{ire}[n]  & 
       V_{iim}[n]  &
       V_{mre}[n]  & 
       V_{mim}[n]  &
       V_{tre}[n]  & 
       V_{tim}[n]  &
       I_{SE1re}[n] &
       I_{SE1im}[n] &
       I_{SE2re}[n] &
       I_{SE2im}[n]  
       \end{array}\end{bmatrix}^{'}$}} 
       \label{IPFC_Vector}
\end{equation}
\begin{equation}
\resizebox{0.9\hsize}{!}{%
\rotatebox{0}{$[B^{IP}] = \begin{bmatrix} \begin{array}{cccccccccccccccc}
\Re[\Gamma_{\mathcal{F}}] &
                   \Im[\Gamma_{\mathcal{F}}] &
                      \Gamma_{\mathcal{G}} &
                      \Gamma_{\mathcal{GV}}  &
				   \Re[\Gamma_{\mathcal{L}}] & 
                   \Im[\Gamma_{\mathcal{L}}] & 
                   \Re[\Gamma^{IP}_{i}] & 
                   \Im[\Gamma^{IP}_{i}] & 
                   \Re[\Gamma^{IP}_{m}] &
                   \Im[\Gamma^{IP}_{m}] &
                   \Re[\Gamma^{IP}_{t}] &
                   \Im[\Gamma^{IP}_{t}] &
                      \Gamma^{IP}_{PBE}         &
                      \Gamma^{SC}_{Mi}    	 &
                      \Gamma^{SC}_{Mi}     	 &
                      \Gamma^{SC}_{Mi}     	      	  
\end{array}\end{bmatrix}^{'} $}} 
\label{IPFC_RHS}
\end{equation}
\end{figure*}

The incorporation of a IPFC introduces two new complex variables $I_{SE1}$ and $I_{SE2}$ into the system, which adds four unknown variables. Therefore, four supplementary equations are required to find out the unique solution of unknown variables. One equation is based on the real power exchange constraint of the device and other three equations are related to the three quantities which are to be controlled. An IPFC can operate in various control modes as suggested in \cite{vinkovic2009current}. The mathematical formulation of different control modes are similar to the control modes of SSSC, only subscript needs to be changed. 

By comparing the coefficient of $\alpha^{n}$, the general recurrence relationship for $n\geq 1$ can be obtained. Further, the system of linear equations for IPFC $[A^{IP}]_{(\Upsilon\times\Upsilon)}[X^{IP}]_{(\Upsilon\times1)}=[B^{IP}]_{(\Upsilon\times1)}$ has been obtained by shifting all the unknown and known variables to the LHS and RHS respectively and $\Upsilon=2(N+n_{IP})$. The coefficient matrix $[A^{IP}]$, unknown vector $[X^{IP}]$ and known vector $[B^{IP}]$ are presented in (\ref{IPFC_Matrix}), (\ref{IPFC_Vector}) and (\ref{IPFC_RHS}) respectively (shown at the bottom of this page). The control modes presented in (\ref{SSSC_Modes}) have been adopted with necessary subscript changes and investigated in numerical studies. The entries of known vector $[B^{IP}]$ are presented in Appendix A.

\section{Results and Discussions} \label{results}
The proposed models of the stated devices and their multi-control capabilities have been tested on IEEE 30 and 118-bus test systems \cite{zimmerman2010matpower}. But for demonstration purpose, only selected numerical results on IEEE 118-bus test system have been presented in this paper. In this paper, the impedances of series coupling transformers are assumed to be $0.01+j0.01$ \textit{p.u.} To validate the proposed models, investigation has been done by incorporating the SSSC or IPFC at different locations and by changing the target values of quantities. All the quantities are in \textit{p.u.} and the base power being is 100 MVA. The proposed FFHE based models have been tested and examined for all the control modes as discussed in Section \ref{PM}. The proposed FFHE and NR based models were modelled in MATLAB environment and simulated on Intel(R) Core(TM) i3-4150 CPU 3.50 GHz processor with 4-GB RAM. For all the investigations, the mismatch tolerance of $10^{-8}$ is considered for maximum bus power mismatch. The percentage reduction in mismatch and runtime are calculated using (\ref{Error}) and (\ref{Time}) respectively.
\begin{small}
\begin{equation}
\% \vartriangle E = \frac{|log10 (max(\vartriangle S_{FFHE}))-log10(max(\vartriangle S_{NR}))|}{|log10 (max(\vartriangle S_{NR}))|}\times 100
\label{Error}
\end{equation}
\end{small}
\begin{equation}
\% \vartriangle T = \frac{T_{NR}-T_{FFHE}}{T_{NR}}\times 100
\label{Time}
\end{equation}
where, $\vartriangle S_{FFHE}$ and $\vartriangle S_{NR}$ are the bus power mismatch obtained using FFHE and NR method respectively; and $T_{NR}$ and $T_{FFHE}$ are the runtime of NR and FFHE method to converge respectively.

Note that the proposed embedding may not guarantee to satisfactory performance for all initial guesses of the variables. The selection of promising initial guess still remains an issue. But \cite{chiang2017novel} suggest the integration of FFHE method with traditional iterative load-flow methods as it takes lesser number of terms to converge as compared to basic HELM. The values of unknown variables obtained after 3 iterations of NR method have been used as initial guesses for the FFHE load-flow method throughout this paper. The numerical results of IEEE 118-bus test system in the absence of FACTS devices are presented in Table \ref{Without_table}. 

\begin{table}[width=.7\linewidth,cols=2,pos=t]
\caption{Selected numerical results of IEEE 118-bus test system without any FACTS device}\label{Without_table}
\begin{tabular*}{\tblwidth}{@{}LL@{}}
\toprule 
$V_{49}=1.0250\angle -8.97\degree$    & $S_{49-50}=0.5367+j0.1343$ \\
$V_{100}=1.0170\angle -1.91\degree$   & $S_{49-51}=0.6664+j0.2044$ \\
$V_{101}=0.9928\angle -0.35\degree$   & $S_{100-104}=0.5642+j0.1062$ \\
$S_{101-102}=-0.0535+j0.0245$         & $S_{100-106}=0.6058+j0.0909$ \\
\bottomrule
\end{tabular*}
\end{table}
\begin{table}[t]
\centering
\caption{Numerical results when SSSC in the IEEE 118-bus test system (location: 49-50)}
\resizebox{0.98\textwidth}{!}{
\begin{tabular}{llcccccc}
\toprule
\multicolumn{2}{c}{Cases}  & Case 1 & Case 2 & Case 3  & Case 4  & Case 5 & Case 6 \\ 
\hline 
\multicolumn{2}{c}{Specified Parameters}  & $P^{SP}_{49-50}=0.75$ & $Q^{SP}_{49-50}=0$ & $Q^{SP}_{SE}=0.3$ & $V^{SP}_{49}=1$ & $V^{SP}_{SE}=0.2$ & $X^{SP}_{eq(SE)}=-0.2$ \\
\hline
\multirow{6}{*}{\rotatebox{90}{Power-flow }}\multirow{5}{*}{\rotatebox{90}{results}} & $V_{49}$ & $1.0385\angle -9.38\degree$ &  $1.0438\angle -8.88\degree$ & $1.0101\angle-9.20\degree$ & $\textbf{1}\angle-9.24\degree$ & $1.0060\angle-9.03\degree$ & $1.0161\angle-8.59\degree$ \\

& $V_{SE}$ & $0.0873\angle 72.75\degree$ & $0.2162\angle -270.01\degree$ & $0.2702\angle109.27\degree$ & $0.2696\angle112.54\degree$ & $\textbf{0.2}\angle 100.26\degree$ & $0.0705\angle-115.49\degree$ \\

& $I_{SE}$ & $0.7251\angle-17.25\degree$ & $0$ & $1.1103\angle19.27\degree$ & $1.1113\angle22.54\degree$ &  $0.9617\angle 10.26\degree$ & $0.3524\angle-25.49\degree$ \\

& $S_{49-50}$ & $\textbf{0.75}+j0.1036$ & $0$ & $0.9859-j0.5346$ & $0.9447-j0.5853$ & $0.9132-j0.3196$ & $0.3426+j0.1042$ \\

& $S_{SE}$ & $0+j0.0637$  & $0-\textbf{j0}$ & $0+\textbf{j0.3}$ & $0+j0.2996$ & $0+j0.1923$ & $0-j0.1042$ \\

& $X_{eq(SE)}$ & $0.1198$  & $inf$ & $0.2434$ & $0.2426$ & $0.2080$ & $-\textbf{0.2}$ \\

\hline
\multicolumn{2}{c}{$\% \vartriangle E$} & $26.42$ & $16.21$ & $22.22$ & $21.56$ & $19.69$  & $21.32$\\
\hline
\multicolumn{2}{c}{$\%  T$}     & $8.02$ & $7.14$ & $8.56$ & $6.25$ & $ 5.83$ & $6.34$\\
\bottomrule 
\end{tabular}} 
\label{SSSC_table1}
\end{table}

Now to investigate the proposed embedding for SSSC and its six different controlling modes, tests have been considered for different locations and specified references. For examining cases 1-6, a SSSC is incorporated in the transmission line between buses 49 and 50 at the location of bus 49; and the results are presented in Table \ref{SSSC_table1}. Specified values of quantities, selected power-flow results, SSSC's parameters, percentage reduction in error and run-time are also presented in Table \ref{SSSC_table1}. The calculated values of specified parameters for different cases are shown in bold letters. 
From Table \ref{Without_table}, it can be observed that the normal active and reactive power-flow between the buses 49 and 50 are 0.5367 \textit{p.u.} and 0.1343 \textit{p.u.} respectively. In case 1, the desired value of the active power-flow through the transmission is set to 0.75 \textit{p.u.}, which is about 40\% higher than the natural power-flow. To increase the real power-flow from 0.5367 to 0.75 \textit{p.u.}, SSSC injects the voltage of 0.0873 \textit{p.u.} The FFHE based model reduces error 26.42\% more as compared to NR based model and it takes 8.02\% less runtime.

In case 2, the desired value of the reactive power-flow through the transmission line was set to 0 \textit{p.u.} and from Table \ref{SSSC_table1}, it can be verified that the reactive power-flow through the transmission line is equal to the specified control reference value. In the next case, the reactive power injected by the SSSC was selected as a control variable. Investigation of case 3 verifies that the injected reactive power is equal to the target value of 0.3 \textit{p.u.}.

As discussed earlier, a SSSC can also regulate the magnitude of bus voltage to the desired value. From Table \ref{SSSC_table1}, it can be observed that the SSSC is able to regulate the voltage of bus 49 to the specified reference 1 \textit{p.u.} But in this case, the convergence characteristic is poor. The probable reason for this is the poor coupling between the bus voltage magnitude and series compensation (generally, shunt-connected FACTS devices such as STATCOM and SVC are used for bus voltage magnitude control). In case 5, the target value of the magnitude of SSSC’s injected voltage is set to 0.2 \textit{p.u.} and it can be verified that the proposed model is able to control the output voltage of SSSC to a specified value. This mode is the logical choice as it directly represents the SSSC’s limitations. As discussed in Section \ref{PM}, a SSSC can also operate like an imaginary controllable impedance and it can be verified from results of case 6. In this case, a SSSC regulates its output voltage phasor in such a manner that the SSSC’s imaginary equivalent reactance is equal to the specified reference $-0.2$ \textit{p.u.}

The efficacy of the proposed model was further validated by inserting a SSSC at bus 101 in the transmission line 101-102. From Table \ref{SSSC_table2}, it can be clearly observed that all the target values have been achieved when the limit of injected voltage was not considered. The decrement in error ranges from 18.46\% to 29.45\% and reduction in runtime ranges from 5.15\% to 7.63\%. Therefore, it can be concluded that the suggested NR-assisted FFHE method based SSSC model converges and performs faster.

\begin{table}[pos=b]
\centering
\caption{Numerical results when SSSC in the IEEE 118-bus test system (location: 101-102)}
\resizebox{0.97\textwidth}{!}{
\begin{tabular}{llcccccc}
\toprule 
&  & Case 1 & Case 2 & Case 3  & Case 4  & Case 5 & Case 6 \\ 
\hline 
\multicolumn{2}{c}{Specified Parameters}  & $P^{SP}_{101-102}=0.9$ & $Q^{SP}_{101-102}=0$ & $Q^{SP}_{SE}=0.3$ & $V^{SP}_{49}=0.9$ & $V^{SP}_{SE}=0.1$ & $X^{SP}_{eq(SE)}=0.1$ \\
\hline
\multirow{6}{*}{\rotatebox{90}{Power-flow }}\multirow{5}{*}{\rotatebox{90}{results}}  & $V_{101}$ & $0.9208\angle -12.96\degree$ &  $0.9936\angle -3.98\degree$ & $0.9551\angle-10.59\degree$ & $\textbf{0.9}\angle3.4903\degree$ & $0.9749\angle 1.49\degree$  & $0.9845\angle0.0522\degree$ \\

& $V_{SE}$ & $0.5114\angle 63.82\degree$ & $0.1518\angle 89.30\degree$ & $0.4161\angle71.67\degree$ &  $0.3009\angle-39.07\degree$ & $\textbf{0.1}\angle-64.33\degree$ & $0.0522\angle-69.97\degree$ \\

& $I_{SE}$ & $1.0040\angle-26.18\degree$ & $0$ & $0.7209\angle-18.3297\degree$ & $1.1335\angle-129.07\degree$ & 
$0.6406\angle-154.33 \degree$ & $0.5218\angle-159.97\degree$ \\

& $S_{101-102}$ & $\textbf{0.9}+j0.2115$ & $0$ & $0.6823+j0.0927$ & $-0.69+j0.7514$ & $-0.5697+j0.2557$ & $-0.4844+j0.1709$ \\

& $S_{SE}$ & $0+j0.5134$ & $0-\textbf{j0}$ & $0+\textbf{j0.3}$ & $0+j0.3411$ & $0+j0.0641$ & $0+j0.0272$ \\

& $X_{eq(SE)}$ & $0.5093$ & $inf$ & $0.5773$ & $0.2655$ & $0.1561$ & $\textbf{0.1}$ \\

\hline
\multicolumn{2}{c}{$\% \vartriangle E$} & $29.45$ & $18.46$ & $28.32$ & $19.72$ & $27.05$ & $21.76$\\
\hline
\multicolumn{2}{c}{$\%  T$}             & $7.63$ & $7.21 $ & $8.49$ & $6.52$ & $5.15$ & $5.56$\\
\bottomrule 
\end{tabular}} 
\label{SSSC_table2}
\end{table}

\begin{table}[pos=b]
\caption{Numerical results when IPFC in the IEEE 118-bus test system (location: $49^{50}_{51}$)}
\resizebox{0.5\textwidth}{!}{
\begin{tabular}{llcc}
\hline 
& & Case 1 & Case 2  \\ 
\toprule
\multicolumn{2}{l}{\multirow{3}{1cm}{Specified Parameters}}  & $P^{SP}_{49-50}=0.75$ & $P^{SP}_{49-50}=0.75$ \\
							& 	& 				$P^{SP}_{49-51}=0.75$ & $Q^{SP}_{49-50}=0.01$ \\
							& 	& 				$Q^{SP}_{49-51}=0.03$ &   $Q^{SP}_{49-51}=-0.03$\\

\hline
\multirow{11}{*}{\rotatebox{90}{Power-flow results}} & $V_{49}$ & $1.0613\angle-9.77\degree$ & $1.0427\angle-10.32\degree$ \\

& $V_{SE1}$ & $0.1149\angle19.47\degree$ & $0.1487\angle 85.88\degree$ \\

& $V_{SE2}$ & $0.0886\angle 125.05\degree$ & $0.2596\angle 78.74\degree$ \\

& $I_{SE1}$ & $0.8347\angle-41.93\degree$  & $0.7193\angle-11.08\degree$ \\

& $I_{SE2}$ & $0.7072\angle-12.06\degree$  & $1.2936\angle-9.05\degree$ \\

& $S_{49-50}$ & $\textbf{0.75}+j0.4716$  & $\textbf{0.75+j0.01}$  \\

& $S_{49-51}$ & $\textbf{0.75+j0.03}$  & $1.3485\textbf{-j0.03}$ \\

& $S_{SE1}$  & $0.0459+j0.00842$ & $-0.0130+j0.1062$ \\

& $S_{SE2}$ & $-0.0459+j0.0426$ & $0.0130+j0.3355$ \\

& $X_{eq(SE1)}$ & $0.1208$ & $0.2052$ \\

& $X_{eq(SE2)}$ & $0.0852$ & $0.2005$ \\

\hline
\multicolumn{2}{c}{$\% \vartriangle E$} & $24.53$ & $21.46$ \\
\hline
\multicolumn{2}{c}{$\%  T$}             & $4.41$ & $6.56$ \\
\bottomrule 
\end{tabular}}
\label{IPFC_table1}
\end{table}

\begin{table}[!b]
\centering
\caption{Numerical results when IPFC in the IEEE 118-bus test system (location: $100^{104}_{106}$)}
\begin{tabular}{llcc}
\toprule 
&  & Case 1 & Case 2  \\ 
\hline 
\multicolumn{2}{l}{\multirow{3}{1cm}{Specified Parameters}} & $P^{SP}_{100-104}=0.8$ & $P^{SP}_{100-104}=0.9$ \\
              &  & $P^{SP}_{100-106}=1$ & $Q^{SP}_{100-104}=0$ \\
               & & $Q^{SP}_{100-106}=0$ &   $Q^{SP}_{100-106}=0$\\
\hline
\multirow{11}{*}{\rotatebox{90}{Power-flow results}} & $V_{100}$ & 1.0184$\angle-$8.47$\degree$ & 1.0170$\angle$ -5.04$\degree$ \\

& $V_{SE1}$ & $0.0742\angle 150.96\degree$ &   $0.0516\angle 84.55\degree$ \\

& $V_{SE2}$ & $0.2612\angle 69.90\degree$ & $0.0223\angle 86.66\degree$ \\

& $I_{SE1}$ & $0.7933\angle-0.42\degree$ &  $0.8850\angle-5.04\degree$ \\

& $I_{SE2}$ & $0.9819\angle-8.47\degree$ & $0.4864\angle-5.04\degree$ \\

& $S_{100-104}$ & $\textbf{0.8}-j0.1132$ & $\textbf{0.9+j0}$  \\

& $S_{100-106}$ & $\textbf{1-j0}$ & $0.4947+\textbf{j0}$ \\

& $S_{SE1}$ & $-0.0517+j0.0282$ & $0.0003+j0.0456$ \\

& $S_{SE2}$ & $0.0517+j0.2512$ & $-0.0003+j0.0109$ \\

& $X_{eq(SE1)}$ & $0.0448$ & $0.0583$ \\

& $X_{eq(SE2)}$ & $0.2605$ & $0.0459$ \\

\hline
\multicolumn{2}{c}{$\% \vartriangle E$} & $21.69$ & $23.64$ \\
\hline
\multicolumn{2}{c}{$\%  T$}             & $3.02$ & $ 4.86$ \\
\bottomrule
\end{tabular} 
\label{IPFC_table2}
\end{table}

To validate the proposed embedding for an IPFC, numerical studies has been carried out for various location of an IPFC along-with different controlling modes. Some selected results on IEEE 118-bus test system for two cases have been presented in Table \ref{IPFC_table1} and \ref{IPFC_table2} when a generic IPFC consisting of two branches is connected in the system. The reference quantities being the real and reactive power-flows through the transmission line, as these are the practice for the series devices in the literature. 
For examining both the cases an IPFC of two branches (branch 1 is connected between buses 49 and 50; and branch 2 is connected between buses 49-51) is integrated into the system and the results are presented in Table \ref{IPFC_table1}. In case 1, the active power flowing through the lines 49-50 and 49-51 were set to 0.75 \textit{p.u.}, which is higher than the natural power flow; and the reactive power-flow through the line 49-51 was set to 0.03 \textit{p.u.} In case 2, the active power flowing through the line 49-50 was set to 0.75 \textit{p.u.}, and the reactive power-flow through the lines 49-50 and 49-51 were set to 0.01 \textit{p.u.} and $-$0.03 \textit{p.u.} respectively. From Table \ref{IPFC_table1}, it can be observed that the specified reference values in both cases are met. From this table, it can also be observed that the active power exchanged between two series converters through common DC link is 0.0459 \textit{p.u.} and 0.0130 $p.u.$ in case 1 and 2 respectively. Table \ref{IPFC_table2} presents the results for two cases when the IPFC is assumed to be connected between buses 100-104 and 100-106. From Tables \ref{IPFC_table1} and \ref{IPFC_table2}, it can be observed that the error reduction is higher and run-time is lesser for FFHE based model. Note that in all the cases, convergence has been achieved; and the proposed models exhibited very good convergence and also converge at faster rate as compared to models based on standard NR method.

The introduced embedding system has also been investigated with device limit constraints. To demonstrate the handling of constraints, one case from Table \ref{SSSC_table2} was chosen. The maximum value of injected voltage by SSSC was selected as 0.3 \textit{p.u.} From Table \ref{SSSC_table2}, it can be observed that when the active power-flow through transmission line 101-102 is chosen as 0.9 \textit{p.u.} and no limits were set on the injected voltage, the SSSC is able to control the active power-flow to the specified reference. But, when the limits are imposed, SSSC is not able to control the same because higher $V_{SE}$ is required to maintain the same. Therefore, whenever the limit constraints are violated, existing control mode is relaxed; and SSSC will acts as a constant voltage source (\textit{i.e.} injected voltage magnitude mode is triggered) and the specified reference value of injected voltage magnitude is set equal to the value of violated limit. From Table \ref{Limits_table}, it can be observed that the active-power flow through the line 101-102 is not equal to the specified reference value but it is equal to 0.3860 \textit{p.u.} while the injected voltage magnitude is maintained to 0.3 \textit{p.u.} Therefore, it can be concluded that the proposed FFHE based models are also suitable when limit constraints of the devices are imposed.

\begin{table}[t]
\centering
\caption{Power-flow results when device limit constraints are imposed  }
\begin{tabular}{cccl}
\hline 
Device  & Specified Parameter (\textit{p.u.}) & Limit Violated & Power-flow results \\
\hline 
\multirow{8}{*}{SSSC}   & \multirow{8}{*}{$P^{SP}_{101-102}=0.9$} & \multirow{8}{*}{Yes} & $V_{101}=0.9891\angle-7.68\degree$\\
                                         &                   &     & $V_{SE}=\textbf{0.3}\angle82\degree$\\
                                         &                   &     & $I_{SE}=0.3902\angle-8\degree$\\
                                         &                   &     & $S_{101-102}=\textcolor{blue}{0.3860}+j0.0021$\\
                                         &                   &     & $S_{SE}=0+j0.1171$\\
                                         &                   &     & $X_{eq(SE)}=0.7687$\\
                                         &                   &     & $\% \vartriangle E=21.99$\\
                                         &                   &     & $\%  T=4.31$\\
\hline
\end{tabular}
\label{Limits_table}
\end{table}

The NR and FFHE based model of SSSC and IPFC have been tested on same platform for comparison, although the derived inferences may not be general due to various differences in simulation structure, programming skills \textit{etc.} It is observed that the percentage reduction in error and runtime are different for each case. Because there is no particular choice of the initial guess for the proposed models to ensure fast convergence in all cases. Briefly, the runtime and rate of convergence of the proposed models will vary for different initial guess. Although, the performance of proposed FFHE based model of SSSC and IPFC are sensitive to the initial guess, but the proposed models perform better than basic HELM based models. Moreover, for most of cases, FFHE based models outperforms NR based models in context of error reduction. A common conclusion for the stated devices and their controlling modes is that the proposed embedding for the devices are reliable and have good convergence characteristics. Further, the proposed formulation reduces the mismatch error at faster rate and therefore, converges slightly faster than the NR method based FACTS devices models. Therefore, the proposed FFHE based models offers a good alternative to the NR based models when promising initial guesses are available.  

\section{Conclusion} \label{conclusions}
The purpose of this research work is to develop fast and flexible HELM models of SSSC and IPFC. To do the same, the power balance equations and devices' controlling modes are embedded with a complex variable $\alpha$ in such a manner that the embedded equations satisfy the requirement of embedding. Afterwards, the unknown variables were represented by power series and recursive formulas have been obtained for $n\geq1$. Lastly, at $\alpha=1$ numerical values of unknown variables are calculated using determinant method. The proposed embedded system of stated devices provides flexibility because any state can serve as an initial guess.  

From the numerical results, it is observed that the results of FFHE based models are similar to results obtained by NR based model of devices in perspective of the final calculated values of devices parameters, system variables and operational bounds. The comparison of runtime for FFHE based models with NR based models showed that later one took more time. Although the performance of FFHE based models is sensitive to the initial guess but the NR assisted FFHE based models outperform the NR based models in error reduction. The numerous results showed that the FFHE based models represents a step forward compared to the NR based models. So, the proposed models may be very useful at the planning stage of the power system. Therefore, the proposed FFHE based models offers a good alternative to the NR based models when promising initial guesses are available.

In this paper, no sophisticated technique is used during implementation of FFHE based models, therefore, possibility of reducing runtime further is still large via parallel computing, code optimization \textit{etc.}

\appendix
\section{Entries of Coefficient matrix and Known Vector}
The entries of coefficient matrix $[A^{SC}]$ are as follows:
\begin{equation}
\mu_{ik} = G_{ik}C_{ire} + B_{ik}C_{iim}~~;~ k\neq i 
\end{equation}
\begin{equation}
\xi_{ik} = G_{ik}C_{iim} - B_{ik}C_{ire}~~;~ k\neq i
\end{equation}
\begin{equation}
\mu_{ii} = G_{ii}C_{ire} + B_{ii}C_{iim} + \sum^{N}_{k=1} \big ( G_{ik}C_{kre} - B_{ik} C_{kim} \big )
\end{equation}
\begin{equation}
\xi_{ii} = G_{ii}C_{iim} - B_{ii}C_{ire} + \sum^{N}_{k=1} \big ( G_{ik}C_{kim} + B_{ik} C_{kre} \big )
\end{equation}

\begin{equation}
\mu^{\bigstar}_{ii} = - G_{ii}C_{iim} + B_{ii}C_{ire} + \sum^{N}_{k=1} \big ( G_{ik}C_{kim} + B_{ik} C_{kre} \big )
\end{equation}
\begin{equation}
\xi^{\bigstar}_{ii} = G_{ii}C_{ire} + B_{ii}C_{iim} + \sum^{N}_{k=1} \big ( B_{ik}C_{kim} - G_{ik} C_{kre} \big ) 
\end{equation}

The entries of known vector $[B^{SC}]$ are as follows:
\begin{equation}
\Gamma_{\mathcal{F}} = \eta_{n1}(V^{SP}_{\mathcal{F}}-C_{\mathcal{F}})
\end{equation}
\begin{equation}
\Gamma_{\mathcal{G}} = \frac{\eta_{n1}}{2} \Bigg [ 2P_{\mathcal{G}} - \Re \bigg \{C_{\mathcal{G}}\sum^{N}_{k=1}Y^{*}_{\mathcal{G}k}C^{*}_k \bigg \} \Bigg ] - \Re \Bigg [ \sum^{N}_{k=1}  \sum^{n-1}_{d=1} Y^{*}_{\mathcal{G}k} V^{*}_{\mathcal{G}}[d]V_{k}[n-d] \Bigg ]
\end{equation}
\begin{equation}
\Gamma_{\mathcal{GV}} = \frac{\eta_{n1}}{2} \big [(V^{SP}_{\mathcal{G}})^{2}  - C_{\mathcal{G}}C^{*}_{\mathcal{G}} \big ] - \frac{1}{2}  \sum^{n-1}_{d=1} V_{\mathcal{G}}[d]V^{*}_{\mathcal{G}}[n-d] 
\label{PV_Cons}
\end{equation}
\begin{equation}
\Gamma_{\mathcal{L}} = \eta_{n1} \bigg [ S^{*}_{\mathcal{L}} - C^{*}_{\mathcal{L}} \sum^{N}_{k=1} Y_{\mathcal{L}k}C_{k} \bigg ]
 - \sum^{N}_{k=1}  \sum^{n-1}_{d=1} Y_{\mathcal{L}k} V^{*}_{\mathcal{L}}[d]V_{k}[n-d] 
\end{equation}
\begin{equation}
\Gamma^{SC}_{i} = \eta_{n1} \bigg [ S^{*}_{i} - C^{*}_{i} \sum^{N}_{k=1} Y_{ik}C_{k} - C^{*}_{i}D_{SE} \bigg ] - \sum^{N}_{k=1}  \sum^{n-1}_{d=1} Y_{ik} V^{*}_{i}[d]V_{k}[n-d] -  \sum^{n-1}_{d=1} V^{*}_{i}[d]I_{SE}[n-d] 
\end{equation}
\begin{equation}
\Gamma^{SC}_{m} = \eta_{n1} \bigg [ S^{*}_{m} - C^{*}_{m} \sum^{N}_{k=1} Y_{mk}C_{k} + C^{*}_{m}D_{SE} \bigg ] - \sum^{N}_{k=1}  \sum^{n-1}_{d=1} Y_{mk} V^{*}_{m}[d]V_{k}[n-d] +  \sum^{n-1}_{d=1} V^{*}_{m}[d]I_{SE}[n-d]
  \end{equation}
\begin{equation}
\Gamma^{SC}_{PBE} = - \eta_{n1} \Re \big [ (C_{m}-C_{i})D^{*}_{SE} \big ]  - \Re \Bigg [  \sum^{n-1}_{d=1} \big (V_{m}[d]I^{*}_{SE}[n-d] - V_{i}[d]I^{*}_{SE}[n-d] \big ) \Bigg ]
\end{equation}
\begin{equation}
\Gamma^{SC}_{M1} = \eta_{n1}  \big [P^{SP}_{im} - \Re(C_{i}D^{*}_{SE}) \big ] - \Re \Bigg [  \sum^{n-1}_{d=1} V_{i}[d]I^{*}_{SE}[n-d]  \Bigg ]
\end{equation}
\begin{equation}
\Gamma^{SC}_{M2} = \eta_{n1}  \big [Q^{SP}_{im} - \Im(C_{i}D^{*}_{SE}) \big ] - \Im \Bigg [  \sum^{n-1}_{d=1} V_{i}[d]I^{*}_{SE}[n-d]  \Bigg ]
\end{equation}
\begin{equation}
\Gamma^{SC}_{M3} = \eta_{n1}  \big [Q^{SP}_{SE} - \Im\big \{(C_{m}-C_{i})D^{*}_{SE}\big\} \big ]  - \Im \Bigg [  \sum^{n-1}_{d=1} \bigg \{ V_{m}[d]I^{*}_{SE}[n-d]-V_{i}[d]I^{*}_{SE}[n-d]\bigg \}  \Bigg ]
\end{equation}
$\Gamma^{SC}_{M4}$ can be calculated using (\ref{PV_Cons}) by replacing subscript $\mathcal{G}$ by $i$. As discussed earlier, $F_{SE}(\alpha)$ and $|I_{SE}(\alpha)|$ are the inverse and magnitude of $I_{SE}(\alpha)$, so, the general recurrence relationship between these can be expressed as given in (\ref{inverse}) and (\ref{mag}).
\begin{equation}
F_{SE}[n] = \frac{-1}{D_{SE}}\sum^{n-1}_{d=0}F_{SE}[d]I_{SE}[n-d]
\label{inverse}
\end{equation}
\begin{equation}
\big|I_{SE}[n]\big| = \frac{1}{2\big|D_{SE}\big|}\bigg\{ \sum^{n}_{d=0}I_{SE}[d]I^{*}_{SE}[n-d] 
 - \sum^{n-1}_{d=1}\big|I_{SE}[d]\big|\big|I_{SE}[n-d]\big|\bigg\}
\label{mag}
\end{equation}

\begin{multline}
\Gamma^{SC}_{M5} =  \eta_{n1} \bigg [ V^{SP}_{SE} - \Im \bigg \{ \frac{(C_{m}-C_{i})|D_{SE}|}{D_{SE}}\bigg \} \bigg ] + \Im \bigg [ \frac{(C_{m}-C_{i})}{2D_{SE}|D_{SE}|}  \sum^{n-1}_{d=1} \bigg \{ |I_{SE}[d]||I_{SE}[n-d]|-I_{SE}[d]I^{*}_{SE}[n-d] \bigg \} \bigg]
\\ + \Im \bigg [ \frac{(C_{m}-C_{i})|D_{SE}|}{D^{2}_{SE}} \sum^{n-1}_{d=1} F_{SE}[d]I_{SE}[n-d] \bigg ]  - \Im \bigg [ \sum^{n-1}_{d} \sum^{d}_{l} \bigg \{ V_{m}[l]F_{SE}[d-l]|I_{SE}[n-d]| 
\\- V_{i}[l]F_{SE}[d-l]|I_{SE}[n-d]| \bigg\}\bigg ]
\end{multline}
\begin{equation}
\Gamma^{SC}_{M6} = \eta_{n1}  \Bigg [X^{SP}_{eq(SE)} - \Im\bigg \{\frac{C_{m}-C_{i}}{D_{SE}}\bigg\} \Bigg ] - \Im \Bigg [  \sum^{n-1}_{d=1} \bigg \{ V_{m}[d]F_{SE}[n-d]-V_{i}[d]F_{SE}[n-d]\bigg \}  \Bigg ]
\end{equation}

The entries of known vector $[B^{IP}]$ are as follows:
\begin{multline}
\Gamma^{IP}_{i} = \eta_{n1} \bigg [ S^{*}_{i} - C^{*}_{i} \sum^{N}_{k=1} Y_{ik}C_{k} - C^{*}_{i}(D_{SE1}+D_{SE2}) \bigg ]
 -  \sum^{n-1}_{d=1} \bigg \{ V^{*}_{i}[d]I_{SE1}[n-d] +  V^{*}_{i}[d]I_{SE2}[n-d] \bigg \}
 \\  - \sum^{N}_{k=1}  \sum^{n-1}_{d=1} Y_{ik} V^{*}_{i}[d]V_{k}[n-d]
\end{multline}

\begin{equation}
\Gamma^{IP}_{m} = \eta_{n1} \bigg [ S^{*}_{m} + C^{*}_{m}D_{SE1} - C^{*}_{m} \sum^{N}_{k=1} Y_{mk}C_{k}  \bigg ] - \sum^{N}_{k=1}  \sum^{n-1}_{d=1} Y_{mk} V^{*}_{m}[d]V_{k}[n-d] +  \sum^{n-1}_{d=1} V^{*}_{m}[d]I_{SE1}[n-d]
\end{equation}

\begin{equation}
\Gamma^{IP}_{t} =  \eta_{n1} \bigg [ S^{*}_{t} + C^{*}_{t}D_{SE2} - C^{*}_{t} \sum^{N}_{k=1} Y_{tk}C_{k}  \bigg ] - \sum^{N}_{k=1}  \sum^{n-1}_{d=1} Y_{tk} V^{*}_{t}[d]V_{k}[n-d]  +  \sum^{n-1}_{d=1} V^{*}_{t}[d]I_{SE2}[n-d]
\end{equation}

\begin{multline}
\Gamma^{IP}_{PBE} = - \eta_{n1} \Re \big [ (C_{m}-C_{i})D^{*}_{SE1} + (C_{t}-C_{i})D^{*}_{SE2}\big ] 
- \Re \Bigg [  \sum^{n-1}_{d=1} \big (V_{m}[d]I^{*}_{SE1}[n-d] - V_{i}[d]I^{*}_{SE1}[n-d] 
 \\ + V_{t}[d]I^{*}_{SE2}[n-d] - V_{i}[d]I^{*}_{SE2}[n-d] \big )\Bigg ]
\end{multline}

\bibliographystyle{elsarticle-num}


\bibliography{FFHEREF}


%
%

\end{document}